\documentclass[aps,prb,reprint,superscriptaddress,showpacs,longbibliography,floatfix,footnoteinbib]{revtex4-2}
\usepackage{amsmath,amssymb,graphicx,units}
\usepackage[plainpages=false,pdfpagelabels,colorlinks=true,linkcolor=red,urlcolor=blue,citecolor=blue,pdftitle={Title},pdfauthor={},pdfdisplaydoctitle=true,pdfduplex=DuplexFlipLongEdge]{hyperref}
\usepackage{multirow}
\usepackage{setspace}

\graphicspath{{./figs/}{./}}

\begin{document}

\title{Disorder in interacting quasi-one-dimensional systems: flat and dispersive bands}

\author{Mi-Ji Liang}
\thanks{Those authors contributed equally to this work.}
\affiliation{Key Laboratory of Quantum Theory and Applications of MoE $\&$ Lanzhou Center for Theoretical Physics $\&$ Key Laboratory of Theoretical Physics of Gansu Province, Lanzhou University, Lanzhou, Gansu 730000, China}

\author{Yong-Feng Yang}
\thanks{Those authors contributed equally to this work.}
\affiliation{Key Laboratory of Quantum Theory and Applications of MoE $\&$ Lanzhou Center for Theoretical Physics $\&$ Key Laboratory of Theoretical Physics of Gansu Province, Lanzhou University, Lanzhou, Gansu 730000, China}

\author{Chen Cheng}
\email{chengchen@lzu.edu.cn}
\affiliation{Key Laboratory of Quantum Theory and Applications of MoE $\&$ Lanzhou Center for Theoretical Physics $\&$ Key Laboratory of Theoretical Physics of Gansu Province, Lanzhou University, Lanzhou, Gansu 730000, China}

\author{Rubem Mondaini}
\email{rmondaini@csrc.ac.cn}
\affiliation{Beijing Computational Science Research Center, Beijing 100193, China}

\begin{abstract}
We use the density-matrix renormalization group method to investigate the superconductor-insulator transition (SIT) in disordered quasi-one-dimensional systems. Focusing on the case of an interacting spinful Hamiltonian at quarter-filling, we contrast the differences arising in the SIT when the parent non-interacting model features flat or dispersive bands. Furthermore, we unveil the critical disorder amplitude that triggers insulating behavior by comparing disorder distributions that preserve or not SU(2)-symmetry. While scaling analysis suggests the transition be of a Berezinskii-Kosterlitz-Thouless type for all models (two lattices and two disorder types), only in the flat-band model with Zeeman-like disorder the critical disorder is nonvanishing. In this sense, the flat-band structure does strengthen superconductivity in the presence of attractive interactions. For both flat and dispersive band models, i) in the presence of SU(2)-symmetric random chemical potentials, the disorder-induced transition is from superconductor to insulator of singlet pairs; ii) for the Zeeman-type disorder, the transition is from superconductor to insulator of unpaired fermions. In all cases, our numerical results suggest no intermediate disorder-driven metallic phase. 
\end{abstract}

\maketitle

\section{Introduction}
\label{sec:Intro}

Band dispersion naturally affects the physics of quantum systems. Compared to regular dispersive bands, systems exhibiting flat bands support abundant phenomena such as topological insulating/superconducting physics, various edge states, and exotic superfluid phases. In the noninteracting case, a purely flat band has constant energy as a function of quasimomentum. For a particle loaded in a flat band, the high degeneracy causes it to localize in a compact form within a few sites whose geometry depends on the details of the Hamiltonian~\cite{Maimaiti2017, Rontgen2018, Maimaiti2019, Zhang2020}. Any finite interaction will be much larger than the bandwidth, leading to rich strongly-correlated physics at any value of the interaction strength. 

Among the many interesting aspects of such systems, one of the interests lies in the interplay of the flat band structure, and superconductivity~\cite{Kopnin2011,Peotta2015,Huang2019,Tovmasyan2018,Mondaini2018,Gremaud2021,Chan2022,Chan2022b,Iglovikov2014}. Studies on topological models suggest that the isolated flat bands have much higher superconducting transition temperature~\cite{Kopnin2011,Peotta2015,Huang2019}. In lattice models with flat bands, the preformed pairs dominate transport even above the critical temperature of the transition to a superfluid state~\cite{Tovmasyan2018}. Compared to a standard two-leg fermionic ladder, recent work argues that the Creutz lattice~\cite{Creutz1999} exhibiting a flat dispersion in the non-interacting regime has longer-ranged pairing correlation function, suggesting a more robust pairing and superconductivity~\cite{Mondaini2018}.

One way to further probe whether pairing and superconductivity are enhanced in systems with flat dispersion is to estimate their robustness against disorder. In the absence of interactions,  disorder generically localizes all single-particle eigenstates and induces Anderson localization~\cite{Anderson1958}; in flat-band systems, such localization phenomenon still occurs but with characteristic critical exponents that depend on specific details~\cite{Flach2014, Leykam2017}, including the existence of coupling to dispersive bands~\cite{Leykam2013}. 

At high energies, the interplay  between disorder and interactions can lead to disorder-free flat-band localization at weak disorder, and conventional disorder-induced many-body localization in the strong-disorder regime~\cite{Kuno2020,Danieli2020,He2021,Orito2021,Orito2022}. Turning to the low-energy but yet interacting scenario, sufficient disorder destroys the phase coherence associated with superfluid/superconducting order, leading to a superconductor-insulator transition (SIT)~\footnote{SIT can also occur in clean systems through the tuning of Hamiltonian parameters that govern the single-particle dispersion, e.g., see Refs.~\cite{Mondaini2015,Loh2016,Hazra2020,Jin2022}.}. However, whether the route to insulating behavior proceeds through the direct localization of Cooper pairs~\cite{Feigelman2007} or via the destruction of Cooper pairing then followed by the standard localization of single electrons is still unsettled~\cite{Sacepe2011,Bouadim2011}. Although disorder-induced ground-state transitions have been extensively studied for either spin or bosonic lattices~\cite{Fisher1989,Laflorencie2004,Crepin2011,Meldgin2016}, the transition type and the universality class of disorder-driven SIT in the presence of both charge and spin degrees of freedom remains elusive. 

In this work, we aim to systematically investigate the disorder-induced SIT in a fully interacting setting from the perspective of how robust is the pairing and superconductivity against disorder in systems with either flat or dispersive bands. Additionally, we are also interested in details of the SIT, including its universality class, via proper finite-size scaling from numerically exact calculations of systems with different sizes. Specifically, we focus on the attractive Fermi-Hubbard model on the Creutz lattice with flat dispersion in the noninteracting regime, and carefully examine the pairing and superconductivity via energies, superfluid densities, and correlation functions. We benchmark our results with a regular two-leg ladder with dispersive bands to study how band dispersion affects pairing and superconductivity under the influence of disorder.

The rest of the paper is organized as follows. In Sec.~\ref{sec:model}, we introduce the Hamiltonian on two lattice types with different dispersions, two types of disorder, and also our numerical method. Section~\ref{sec:scaling} is devoted to the finite-size scaling of the superfluid weight, where we discuss the universality class of disorder-induced SIT. In Sec.~\ref{sec:corr}, we further analyze the ground-state phase transition and corresponding phases via correlation functions. The summary of the results is presented in Sec.~\ref{sec:sum}. 

\section{Model and Method}
\label{sec:model}

\begin{figure}[!t] 
 \includegraphics[width=1\columnwidth]{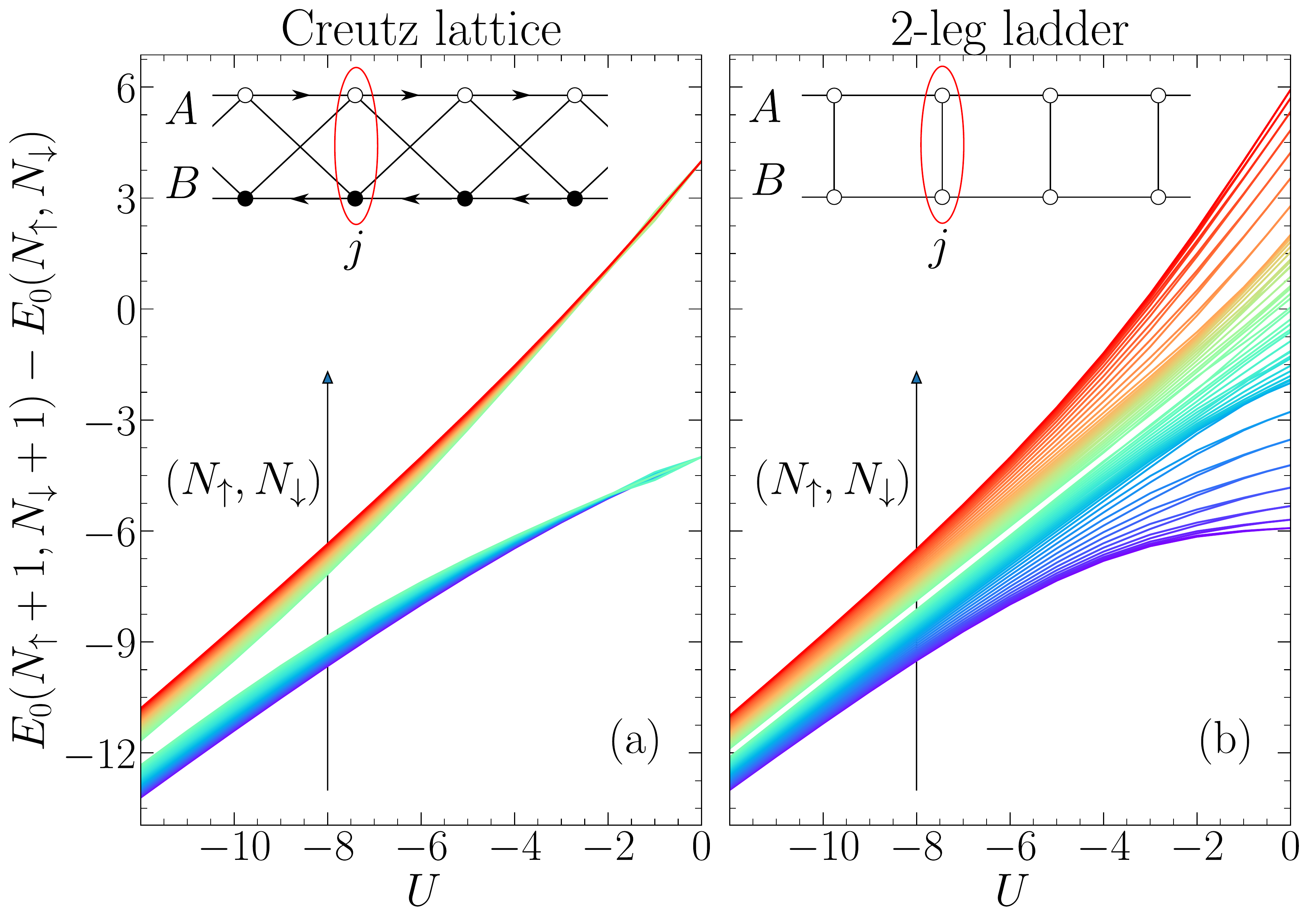}
 \caption{Pair excitation spectrum, $E_0(N_\uparrow+1,N_\downarrow+1)-E_0(N_\uparrow,N_\downarrow)$, in the presence of the negative $U$ for (a) the Creutz lattice and (b) a regular two-leg ladder. Here $E_0(N_\uparrow,N_\downarrow)$ stands for the groundstate energy in a system with $N_\uparrow$ spin-up and $N_\downarrow$ spin-down fermions.  Insets show the lattice geometry: the red ellipse denotes a unit cell labeled by $j$, $A$ and $B$ label two legs. For the Creutz lattice, the arrows depict the sign of the hopping in the intrachain bonds. For the regular two-leg ladder, all hoppings are real without extra phases. Here the results are obtained from numerical calculations of systems with $L=64$ with open boundary conditions.
 }
 \label{fig:band}
\end{figure}

We first consider the attractive Hubbard model on the Creutz lattice described by the Hamiltonian:
\begin{align}
\label{eq:Ham_Creutz}
\nonumber
\hat {\cal H}_C =& -{\rm i}t\sum_{j,\sigma} \left (\hat c^{A\dagger}_{j,\sigma}\hat c^{A}_{j+1,\sigma}
-\hat c^{B\dagger}_{j,\sigma}\hat c^{B}_{j+1,\sigma} - {\rm H.c.}\right )\\
\nonumber
& -t\sum_{j,\sigma} \left ( \hat c^{A\dagger}_{j,\sigma}\hat c^{B}_{j+1,\sigma}
+ \hat c^{B\dagger}_{j,\sigma}\hat c^{A}_{j+1,\sigma} + {\rm H.c.}\right )\\
&  + U \sum_{j,\alpha}
  \hat n^\alpha_{j,\uparrow}\hat n^\alpha_{j,\downarrow},
\end{align}
where $\hat c^{\alpha\dagger}_{j,\sigma}$ ($\hat c^{\alpha}_{j,\sigma}$) creates (annihilates) a fermion with spin $\sigma=\uparrow,\downarrow$ on the $j$-th unit cell with chain index $\alpha=A,B$ [see cartoon in Fig.~\ref{fig:band}(a)];  $\hat n^\alpha_{j,\sigma} = \hat c^{\alpha \dagger }_{j,\sigma} \hat c^{\alpha}_{j,\sigma}$ is the corresponding the number-density operator. For comparison, we also examine the attractive Hubbard model on the regular two-leg ladder:
\begin{align}
\label{eq:Ham_ladder}
\nonumber
\hat {\cal H}_L =& -t\sum_{j,\sigma} \left (\hat c^{A\dagger}_{j,\sigma}\hat c^{A}_{j+1,\sigma}
+ \hat c^{B\dagger}_{j,\sigma}\hat c^{B}_{j+1,\sigma}
+ {\rm H.c.}\right )\\
\nonumber
& -t\sum_{j,\sigma} \left ( \hat c^{A\dagger}_{j,\sigma}\hat c^{B}_{j,\sigma}
+ \hat c^{B\dagger}_{j,\sigma}\hat c^{A}_{j,\sigma} + {\rm H.c.}\right )\\
&  + U \sum_{j,\alpha}
  \hat n^\alpha_{j,\uparrow}\hat n^\alpha_{j,\downarrow}\ ,
\end{align}
where a schematic representation is shown in Fig.~\ref{fig:band}(b). For both models, the linear lattice size is $L$, and the hopping amplitudes are proportional to $t$ (in the Creutz lattice, intrachain hoppings gain a phase, being purely imaginary). The interactions are attractive, $U<0$, wherein fermions with opposite spins form pairs to lower the total energy, further condensing to form a superfluid state, provided the parent state is metallic. 

The Creutz lattice features a flat dispersion in the noninteracting case ($U=0$), and the two-leg ladder has, on the other hand, dispersive bands. An immediate question that arises is how the interaction affects the band structure, which can be inferred by the pair-excitation spectrum (i.e., the chemical potential to introduce a singlet pair) obtained from many-body numerical calculations, as shown in Fig.~\ref{fig:band} (direct comparison of single-particle excitations is given in Appendix~\ref{appendix:A_k}). For the Creutz lattice, there are two highly degenerated bands at $\pm4t$ ($\pm2t$ in the single-particle picture), each with zero bandwidth at $U=0$. Their bandwidth grows in the presence of finite interactions, but the two bands are still relatively narrow. On the contrary, the bandwidth of the two-leg ladder is much larger. Therefore, we assume that the difference in the band structure in the noninteracting case would also affect the pairing and superconductivity in the presence of  interactions. Notice that the model on both lattices at half-filling displays a vanishing superfluid weight ${\cal D}_s$; we focus on filling $\langle \hat n \rangle = \sum_{j,\alpha,\sigma} \langle \hat n_{j,\sigma}^{\alpha}\rangle/2L = 1/4$, to ensure we start from a robust superfluid state. Besides that, in what follows, we fix the interaction strength at $U=-8$ [$t=1$ sets the energy scale], at which ${\cal D}_s$ for the Creutz lattice is close to its maximum in clean cases~\cite{Mondaini2018}. 

The robustness of the superconductivity is estimated by examining the critical disorder to break the pairing coherence and corresponding superconducting state. We first consider the spin-independent random chemical potentials which do not break local singlet pairs:
\begin{align}
\hat {\cal H}_{\mu} = \sum_{i}\mu_{i} \hat n_i,
\label{eq:H_mu}
\end{align}
where $i$ labels a single site and $\mu_{i} \in [-W,W]$ is taken from an uncorrelated, uniform distribution with disorder strength $W$. Alternatively, random Zeeman-like fields introduce another kind of disorder
\begin{align}
\hat {\cal H}_{h} = \sum_{i}h_{i} \hat S^z_i
\label{eq:H_h}
\end{align}
with $h_{i} \in [-W,W]$ and $S^z_i = \hat n_{i,\uparrow} - \hat n_{i,\downarrow}$. The latter breaks SU(2)-symmetry and tends to disassemble pairs (local or not). 

To solve Eqs.~\eqref{eq:Ham_Creutz} and \eqref{eq:Ham_ladder} in the presence of disorder, we numerically employ the density matrix renormalization group (DMRG)~\cite{PhysRevLett.69.2863,PhysRevB.48.10345} method, which is extremely powerful in (quasi-) one-dimensional systems, to obtain the ground state of different lattices, including under disordered settings. The two Hamiltonians have $U(1)$ symmetry with conserved total particle number $N_\sigma =\sum_{i}\langle \hat{n}_{i,\sigma} \rangle$ for spin species $\sigma$ even in the presence of disorder, thus we perform DMRG calculations in the sector with fixed good quantum numbers $N_\sigma$. 
Observables such as superfluid weight, pair binding energy, and correlation functions are computed to characterize the ground state properties. In calculations aiming to obtain the superfluid weight, twisted boundary conditions are used; in the rest of the simulations, we implement open boundary conditions to reduce the computational cost. Up to 2000 DMRG kept states are used in all calculations, and the largest truncation error is about $10^{-6}$. For disordered cases, all observables are obtained from the average over calculations of many disorder samples as indicated in what follows [see Appendix~\ref{appendix:benchmark} for a benchmark against exact results in small systems.]. 

\section{BKT Scaling of Drude weight}
\label{sec:scaling}

\begin{figure}[!t] 
  \includegraphics[width=0.9 \columnwidth]{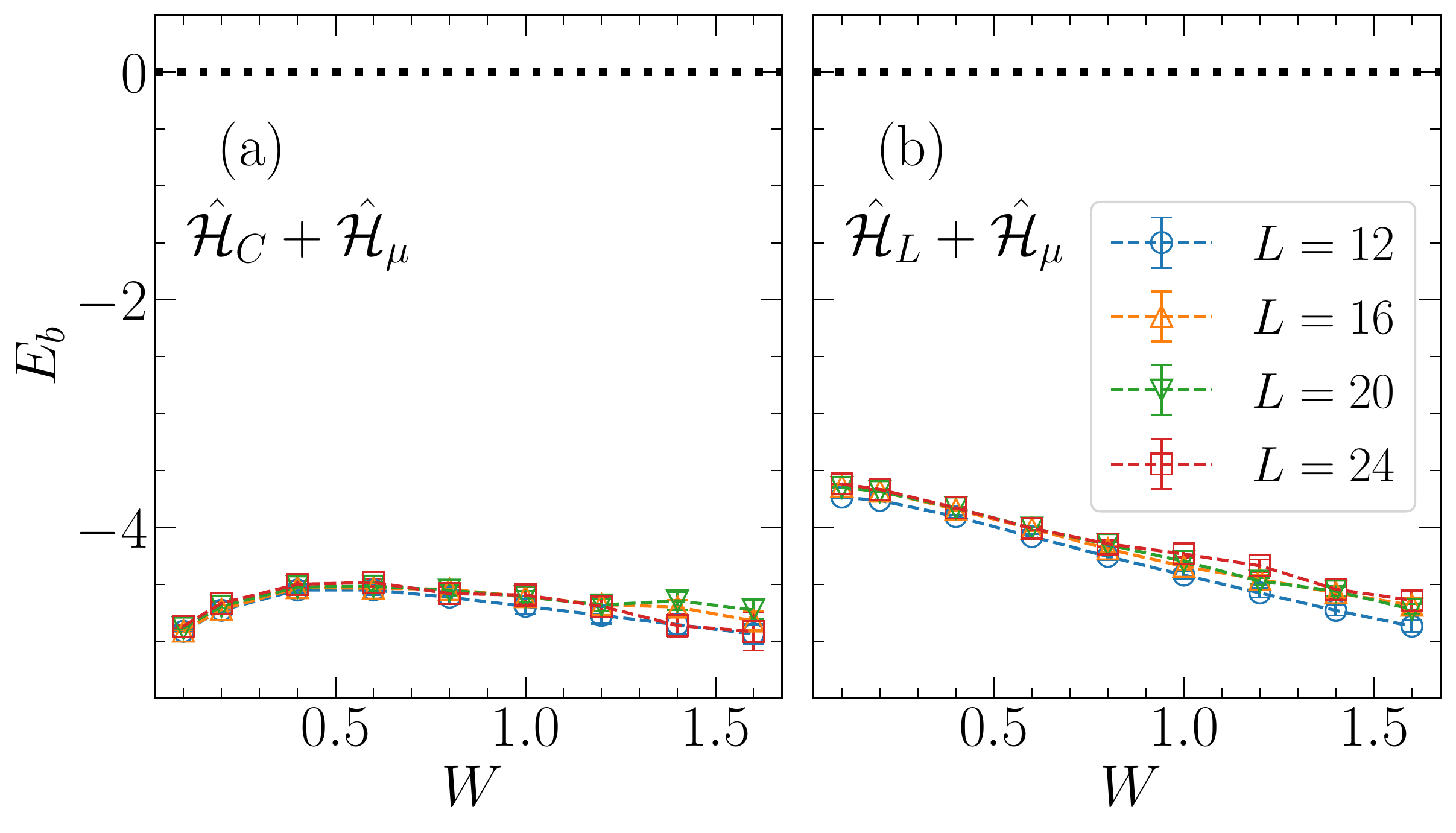}
  \caption{Pairing binding energy $E_b$ as a function of disorder $W$ for (a) the Creutz lattice and (b) regular two-leg ladder with different $L$. Here the disorder is introduced by the random chemical potentials.  }
  \label{fig:Eb_c}
\end{figure}

The pairing of electrons is one of the necessary preconditions of superconductivity, which is a macroscopically coherent state of pairs. We estimate the pair formation via the singlet-pair binding energy
\begin{align}
    E_b \equiv~ &E_0(N_\uparrow+1,N_\downarrow+1) + E_0(N_\uparrow,N_\downarrow) \nonumber \\
    &- 2 E_0(N_\uparrow+1,N_\downarrow).
\end{align}
A negative $E_b$ in the thermodynamic limit denotes that the energy cost for adding two interacting particles (or holes, depending on the filling) with opposite spins is lower than that of two noninteracting ones. As a result, the system exhibits a tendency toward the singlet-pair formation to lower the total energy. We first display the pairing binding energy in the presence of random chemical potentials in Fig.~\ref{fig:Eb_c}. In this case, the binding energy $E_b$ is negative at zero disorder, and remains so with $W>0$, for both the Creutz lattice and regular two-leg ladder. Therefore, from the energetic consideration, fermions in the presence of random chemicals still tend to form pairs, despite the inclusion of disorder.

Whereas useful to characterize singlet pair formation, the binding energy cannot quantify the coherence of such pairs, being thus unable to discern the superconducting state. For that, we examine the superfluid weight, which in one-dimension (1D) is equivalent to the Drude weight~\cite{Kohn1964,Zotos1990,Shastry1990,Fye1991,Scalapino1992,Scalapino1993,Hayward1995}:
\begin{align}
\label{eq:Ds}
{\cal D}_s = \pi L \frac{\partial^2 E_0(\Phi)}{\partial  \Phi^2}\bigg |_{\Phi=0},
\end{align}
where $E_0(\Phi)$ is the ground state in the presence of a threaded magnetic flux $\frac{\hbar c}{e}\Phi$~\cite{Niu1985,Xiao2010}. Such flux is equivalent to the introduction of twisted boundary conditions~\cite{Poilblanc1991} via the replacement $\hat{c}_{j,\sigma} \rightarrow e^{\mathrm{i}\phi j}\hat{c}_{j,\sigma}$, where $\phi = \Phi/L$ is the phase gradient per unit cell. In actual calculations, we use an approximant ${\cal D}_s \approx 2\pi L [E_0(\delta\Phi)-E_0(0)]/(\delta \Phi)^2$~\cite{Laflorencie2004}, choosing $\delta\Phi=\pi/2$ to minimize the numerical error. Thus the superfluid weight is obtained by  
\begin{align}
  \label{eq:approx_Ds}
  {\cal D}_s \approx \frac{8L}{\pi} \left[ E_0(\pi/2) - E_0(0) \right].
\end{align}
The approximation in Eq.~\ref{eq:approx_Ds}, while seemingly crude for such a large $\delta\Phi$, has been numerically confirmed in the clean case, resulting in an absolute error of the order of $10^{-3}$ [see Appendix~\ref{appendix:Ds}].

\begin{figure}[!t] 
  \includegraphics[width=0.9 \columnwidth]{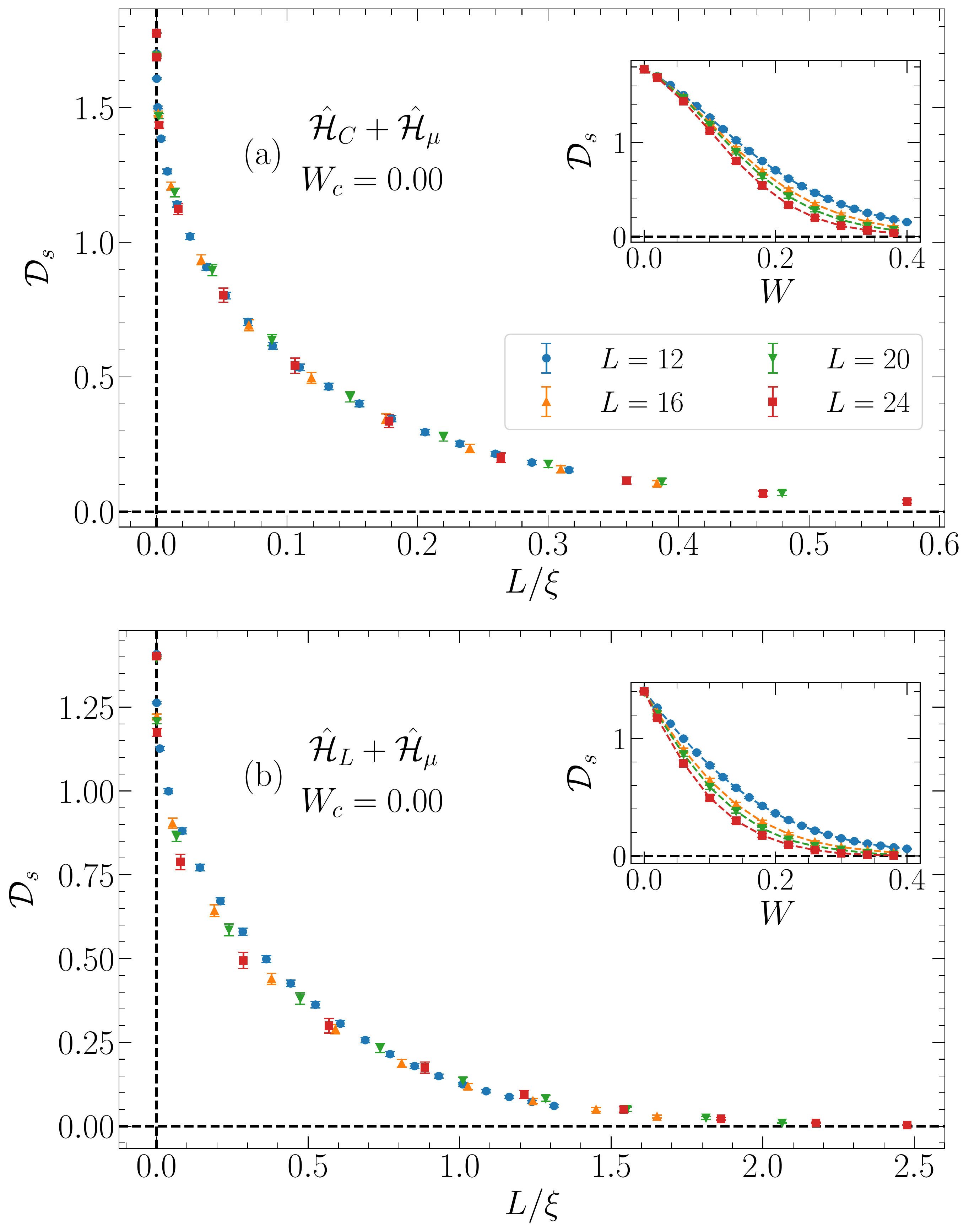}
  \caption{Data collapse of the superfluid weight ${\cal D}_s$ as a function of $L/\xi$ [$-L/\xi$ if $W<W_c$] for (a) the Creutz lattice and (b) regular two-leg ladder with random chemical potentials. Each inset shows ${\cal D}_s$ as a function of the disorder $W$ for different lattice lengths $L$. The optimal parameters $b$ and $W_c$ in Eq.~\ref{eq:xi} are determined by minimizing a cost function of the data collapse [see Appendix~\ref{appendix:Cx} for detailed information]. Numerical results for $L=12$/$16$/$20$/$24$ are obtained by the average over 320/256/160/96 disorder realizations.}
  \label{fig:scaling_C}
\end{figure}

In either 1D or quasi-1D lattices, the quantum phase transition between the Mott insulating phase and the superfluid phase at fixed commensurate lattice filling is known to be of the Berezinskii-Kosterlitz-Thouless (BKT) type~\cite{Fisher1989, Batrouni1990, Scalettar1991, Gerster_2016}. In the case of disordered systems, such scaling form persists for bosonic systems  when transitioning from a superfluid to a Bose glass~\cite{Giamarchi1988,Gerster_2016}. A low-energy effective theory (bosonization) has also been developed for the superfluid-disordered insulating transition in the case of fermionic two-leg ladders~\cite{Orignac1996, Orignac1997, Orignac1999}, suggesting a BKT-type phase transition for the dispersive model. In particular, we notice that the attractive Hubbard model displays the formation of increasingly local Cooper pairs when $|U|\gg 1$, which are mimicked by hardcore bosons in this limit~\cite{Emery1976, Efetov1975, Micnas1990}. Moreover, we recall that in the clean case, the results of the superfluid weight at the strong, attractive interactions we use, $U=-8$, steadily approach the ones for the corresponding bosonic model~\cite{Mondaini2018}, lending further support for the same type of transition similarly occurring here~\footnote{Similar arguments are valid for clean SIT transitions~\cite{Mondaini2015, Jin2022}. There, on the other hand, the transition is of second order [$(d+1)-XY$ universality class], and the fermionic transition shares the same scaling properties of the corresponding mapped bosonic model~\cite{Hen2009, Hen2010}}. Thus assuming a BKT scaling form for the superconductor to the disorder-induced insulator transition (whether the bands are dispersive or not), the disorder-dependent correlation length scales as~\cite{Goremykina2019,Dumitrescu2019,Morningstar2019,Jan2020,Aramthottil2021}  
\begin{align}
    \label{eq:xi}
    \xi = \exp \left\{\frac{b_\pm}{\sqrt{|W-W_c|}}\right\}.
\end{align}
Here $W_c$ is the critical disorder in the thermodynamic limit, and $b_+$ ($b_-$) is a nonuniversal parameter for $W>W_c (W<W_c)$. For numerical convenience, we make the approximation that $b_+ = b_- \equiv b$. Then the critical disorder and the parameter $b$ can be determined by the best data collapse of ${\cal D}_s(L, W)$ as a function of $L/\xi$. 

\begin{figure}[!t] 
  \includegraphics[width=0.9 \columnwidth]{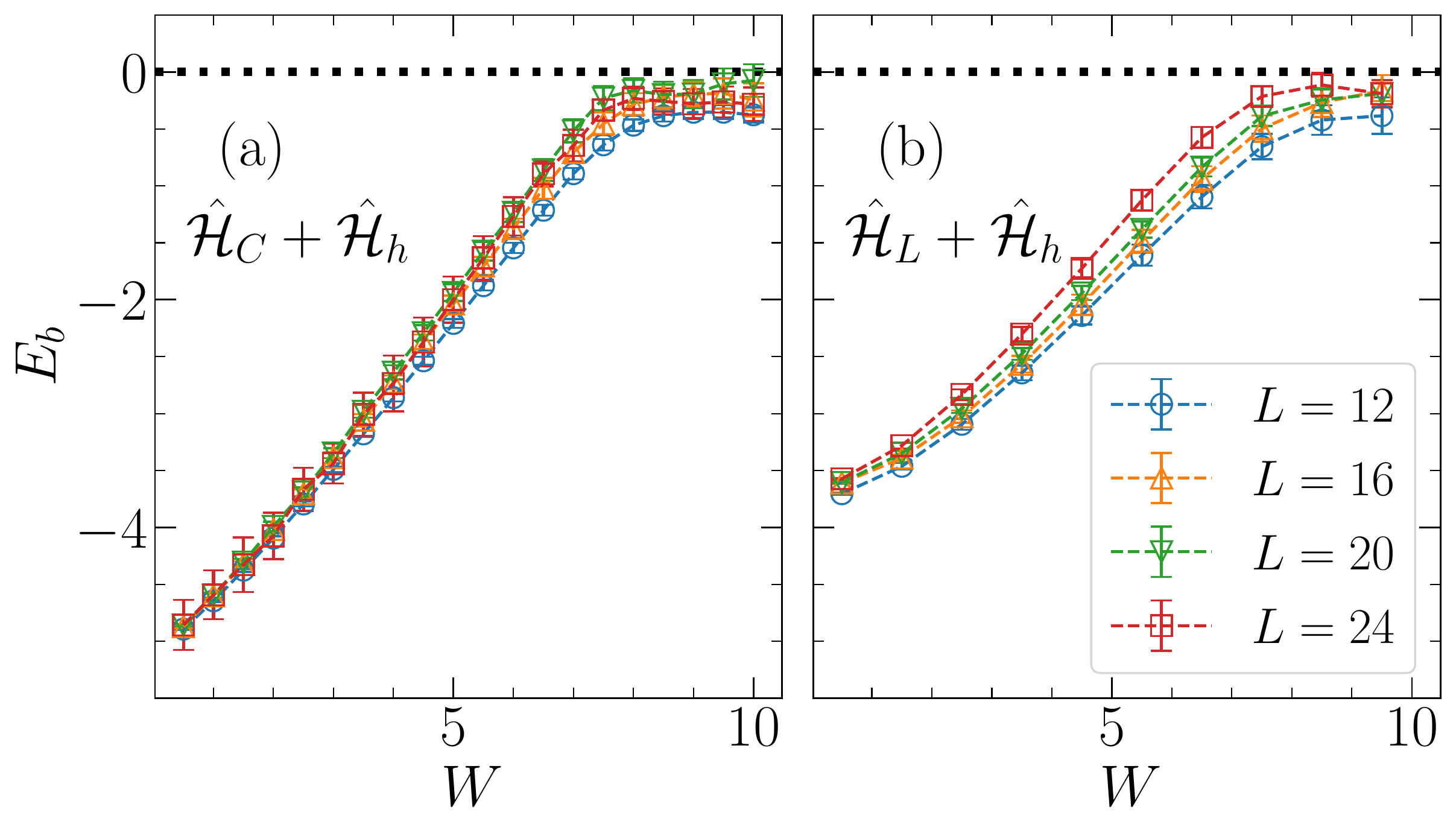}
  \caption{Pairing binding energy $E_b$ as a function of disorder $W$ for (a) the Creutz lattice and (b) regular two-leg ladder with different $L$. Here the disorder is introduced by the random Zeeman field.  }
  \label{fig:Eb_z}
\end{figure}

\begin{figure}[!b] 
  \includegraphics[width=0.9 \columnwidth]{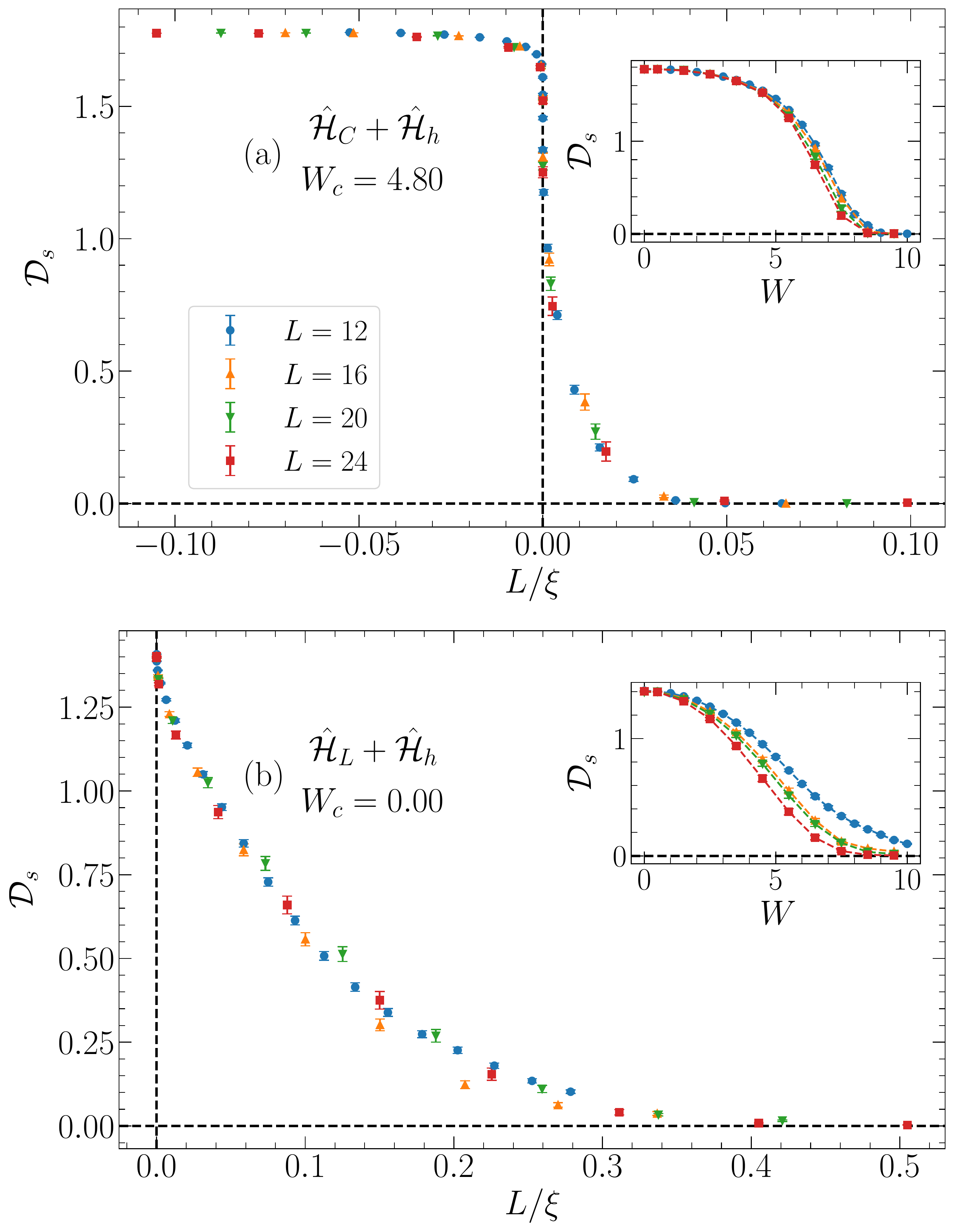}
    \caption{Data collapse of the superfluid weight ${\cal D}_s$ versus $L/\xi$ [$-L/\xi$ if $W<W_c$] for (a) the Creutz lattice and (b) regular two-leg ladder with random Zeeman fields. Other parameters are the same as in Fig.~\ref{fig:scaling_C}.}
  \label{fig:scaling_Z}
\end{figure}

We first consider the disorder introduced by random chemical potentials; the corresponding results of the superfluid weight and its scaling are displayed in Fig.~\ref{fig:scaling_C}. The data collapse of the ${\cal D}_s(L)$ versus $L/\xi$ indicates that the critical disorder $W_c=0^+$ in the thermodynamic limit for both the Creutz lattice and the regular two-leg ladder, within system sizes amenable to our calculations. In this sense, the lattice geometry and band dispersion do not qualitatively affect the (lack of) robustness of the superconductivity against this SU(2)-symmetric disorder. Moreover, the superconducting state is so fragile that an infinitesimal disorder strength destroys the superconductivity. On the other hand, since random chemical potentials do not necessarily break pairs but rather their phase coherence, the system is an (interacting) Anderson insulator of singlet pairs in the disordered phase.

In contrast to the SU(2)-symmetric random chemical potential, the random magnetic field can break the singlet pairs induced by the local attraction $U$. In this case, the introduction of random Zeeman fields with growing disorder strength results in the amplitude of $E_b$ gradually \textit{decreasing} to zero with growing $W$, regardless of the lattice geometry used, as shown in Fig.~\ref{fig:Eb_z}. While this result immediately points out the differences arising from the symmetry-type of disorder used on the pair robustness [see Fig.~\ref{fig:Eb_c} for a comparison], it does not make it clear if pairs are more resilient if contrasting dispersive or dispersionless systems under $\hat {\cal H}_h$. Owing to the lack of a proper and systematic scaling procedure for $E_b$, extracting the critical disorder that breaks pairs only from the binding energy is challenging. Consequently, the results in Fig.~\ref{fig:Eb_z} do not conclude whether the singlet pair is more robust against disorder in the Creutz lattice with flat bands than a regular lattice with dispersive bands.  

To solve this question, we again resort to the superfluid weight, which further probes the phase coherence of the formed pairs. The finite-size scaling of the superfluid weight ${\cal D}_s$ is reported in Fig.~\ref{fig:scaling_Z}. Here, the Creutz lattice and regular two-leg ladder results are qualitatively different under this type of  random Zeeman-like disorder. In contrast to the case with random chemical potentials, the superconducting state survives in the Creutz lattice up until disorder strengths of $W_c\simeq4.8$, as shown in Fig.~\ref{fig:scaling_Z}(a). However, the superconductivity is still fragile and destroyed by an arbitrarily small disorder in the regular two-leg ladder, as shown in Fig.~\ref{fig:scaling_Z}(b). In this sense, the flat dispersion has dramatically enhanced the robustness of the superfluidity, even in the presence of substantial disorder. 


\section{correlation functions}
\label{sec:corr}

\begin{figure}[!b] 
  \includegraphics[width=0.9 \columnwidth]{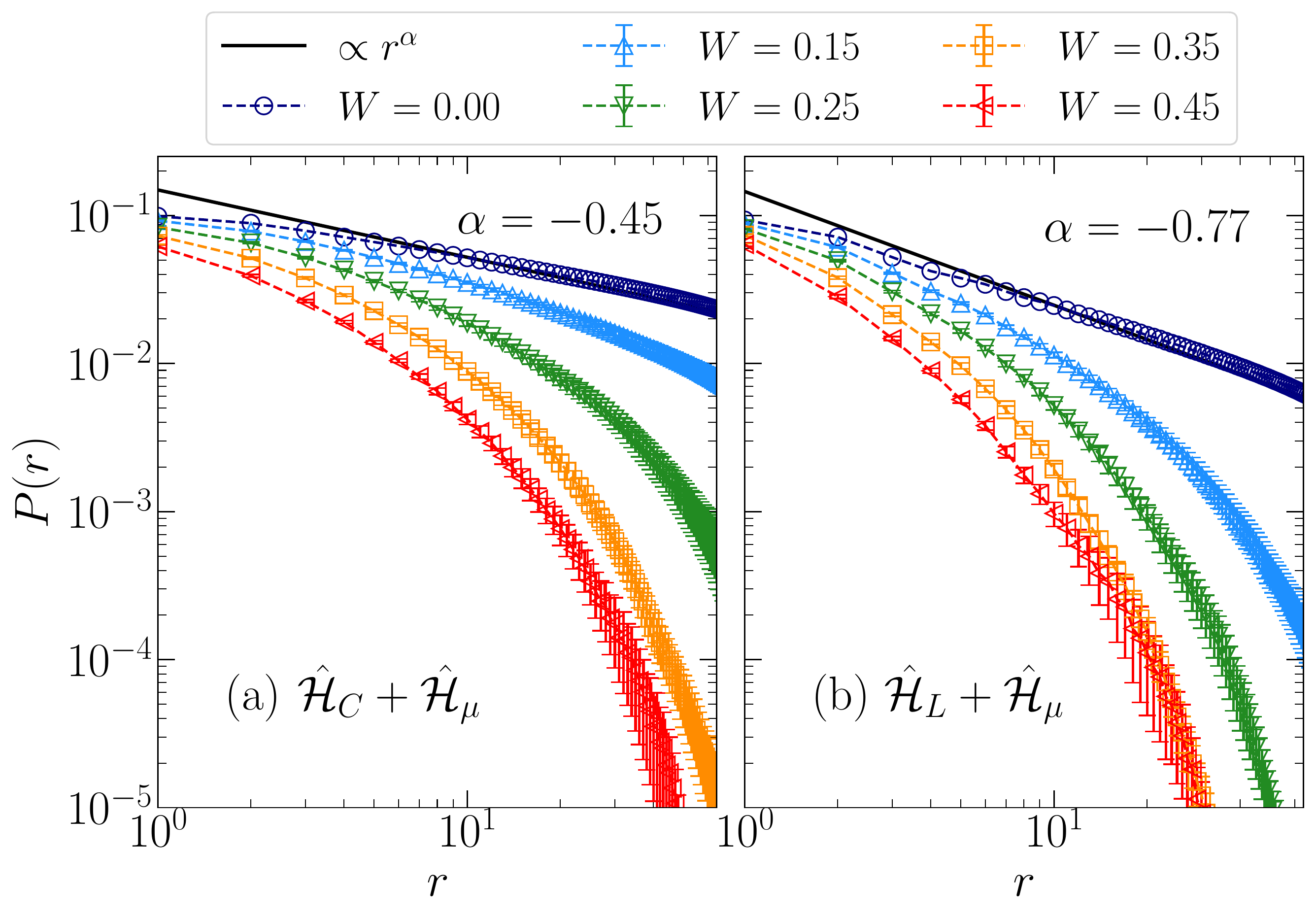}
  \caption{Pairing correlations functions $P(r)$ versus distance $r$ for (a) the Creutz lattice and (b) the regular two-leg ladder. The solid lines give an estimation of a power-law fitting $\propto r^\alpha$ extracted for the clean case that guides the interpretation (note the log-log scale). Here the lattice length $L=128$ and the disorder is introduced via random chemical potentials $\hat {\cal H}_\mu$.}
  \label{fig:corr_Pr_C}
\end{figure}

The scaling of superfluid weight suggests a BKT transition from the superconducting state to the insulating state for both lattice geometries and disorder types. However, the physical characteristics of the disordered phase can be, in principle, different. For example, a large disorder can lead to an Anderson insulator of singlet pairs or even of unpaired fermions. In this section, we calculate the correlation functions to characterize these states further. Specifically, we compute the pairing correlation function,
\begin{align}
G^P_{ij} = \langle \hat \Delta^{A\dagger}_i \hat \Delta^A_j \rangle,
\end{align}
where $\hat \Delta^A_i = \hat c^A_{i,\uparrow}\hat c^A_{i,\downarrow}$ annihilates a local singlet pair on the $i$-th unit cell with chain index $A$, and the single-particle Green's function,
\begin{align}
G^\sigma_{ij} = \langle \hat c^A_{i,\sigma} \hat c^{A\dagger}_{j,\sigma} \rangle.
\end{align}
Computing correlations along one of the chains is sufficient since both lattices have mirror symmetry across the rungs. As mentioned before, we use open boundary conditions for calculations of correlation functions to reduce the computational cost. In this case, for a generic two-point correlation function $X_{ij}$ between sites $i$ and $j$, one can extract the averaged correlation decay as a function of distance as~\cite{Cheng2018,Yang2022}
\begin{align}
\label{eq:Xr}
  X(r) = \frac{1}{\cal N}\sum_{|i-j|=r}X_{ij}\ ,
\end{align}
where $\cal N$ is the total number of pairs $\{i,j\}$ satisfying
$|i-j|=r$. Based on this, we define the average pairing correlation function [single-particle Green's function] as $P(r)$ [$G(r)$].

For the clean case, the system described by the attractive Hubbard model on both lattices features superconductivity, with power-law decaying of pair correlations denoting quasi-long range order. When the disorder is sufficiently strong, the system is in an Anderson-insulating ground state, and the corresponding pairing correlation function decays exponentially. Compared to the Creutz lattice with the flat band, $P(r)$ decays slightly faster in the regular two-leg ladder~\cite{Mondaini2018}. However, the differences in pairing correlations responses to disorder between the two lattices are marginal when random chemical potentials are introduced, as shown in Fig.~\ref{fig:corr_Pr_C}. In this case, $P(r)$ turns to an exponential decay as soon as a minor disorder appears for both lattices, in agreement with the scaling of superfluid weight in Sec.~\ref{sec:scaling}. 

Those results are not fortuitous, as they are precisely aligned with analytic ones from Refs.~\cite{Orignac1996, Orignac1997, Orignac1999}, using bosonization. The Tomonaga-Luttinger exponent of the symmetric charge mode $K_{\rho+}$ describes the stability of the superconducting phase upon inclusion of (non-magnetic) disorder. In particular, when $K_{\rho+} > 3/2 \ (K_{\rho+} < 3/2)$ the superconducting phase is stable (unstable) to $W$. The singlet pair-pair correlations have a general power-law decay of the form, $P(r) \propto r^{-\alpha} = r^{-1/2K_{\rho+}}$~\cite{Giamarchi2004}. In the clean case, the fittings in Fig.~\ref{fig:corr_Pr_C} are compatible with $K_{\rho+} \simeq 1.1$ and 0.65, for the Creutz and the regular ladder, respectively. As a result, they both fall into the regime of unstable superconducting behavior towards including an arbitrarily small disorder, as seen in the results of the superfluid weight [Fig.~\ref{fig:scaling_C}] and predicted in Refs.~\cite{Orignac1996, Orignac1997, Orignac1999} for dispersive bands.

\begin{figure}[!tb] 
  \includegraphics[width=0.9 \columnwidth]{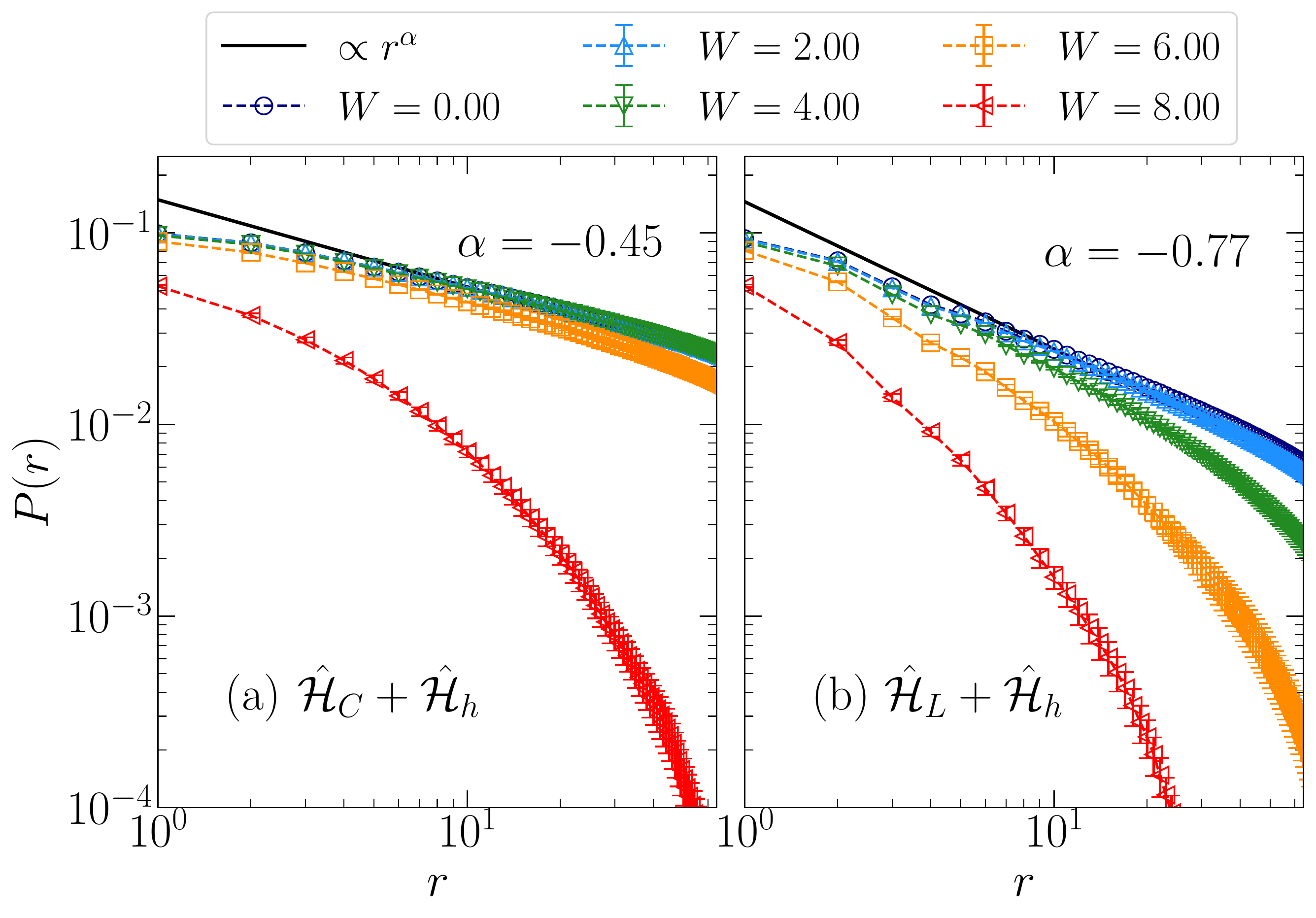}
  \caption{Similar to Fig.~\ref{fig:corr_Pr_C} but for disorder introduced via random Zeeman fields $\hat {\cal H}_h$. The pairing correlation functions $P(r)$ roughly keep their power-law up decay at $W\lesssim6$ for the Creutz lattice (a); the regular ladder (b) shows a much less resilient power-law dependence of $P(r)$. As before, the lattice length is $L=128$ and the solid line gives an estimation of a power-law fitting $\propto r^\alpha$ that guides the interpretation.}
  \label{fig:corr_Pr_Z}
\end{figure}

Differences between the results of the two lattices are much more prominent when random Zeeman fields introduce the disorder, and in this case, no-analytical results stemming from a bosonization analysis exist for such type of magnetic disorder. As shown in Fig.~\ref{fig:corr_Pr_Z}, while $P(r)$ decays exponentially in the presence of a weak disorder strength in the regular two-leg ladder, the pairing correlation function in the Creutz lattice preserves a power-law form even at very large values $W\simeq6$, for the same system size. Notice that although one cannot directly compare the critical disorder from correlation functions at a finite system size with the scaling result of superfluid weights in the thermodynamic limit, these results show qualitatively the same conclusion: the flat-band dispersion dramatically enhances the robustness of the superconductivity against spin-dependent disorder. 

\begin{figure}[!b] 
  \includegraphics[width=0.9 \columnwidth]{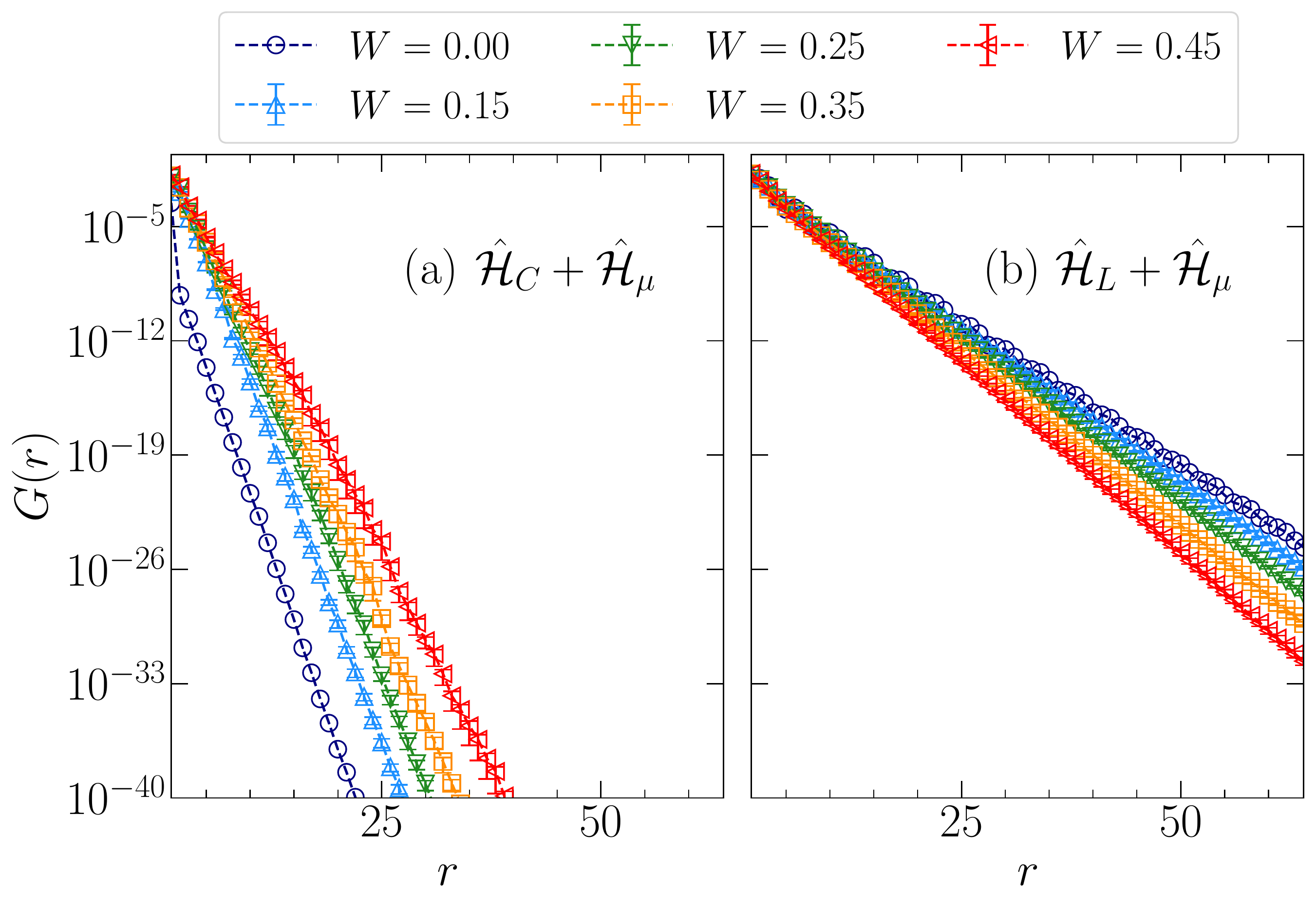}
  \caption{Single-particle Green's functions $G(r)$ versus distance $r$ for (a) the Creutz lattice and (b) the regular two-leg ladder, respectively. Here lattice length $L=128$ and the disorder is introduced by random chemical potentials $\hat{\cal H}_\mu$; note the vertical log-scale.}
  \label{fig:corr_Gr_C}
\end{figure}

\begin{figure}[!t] 
  \includegraphics[width=0.9 \columnwidth]{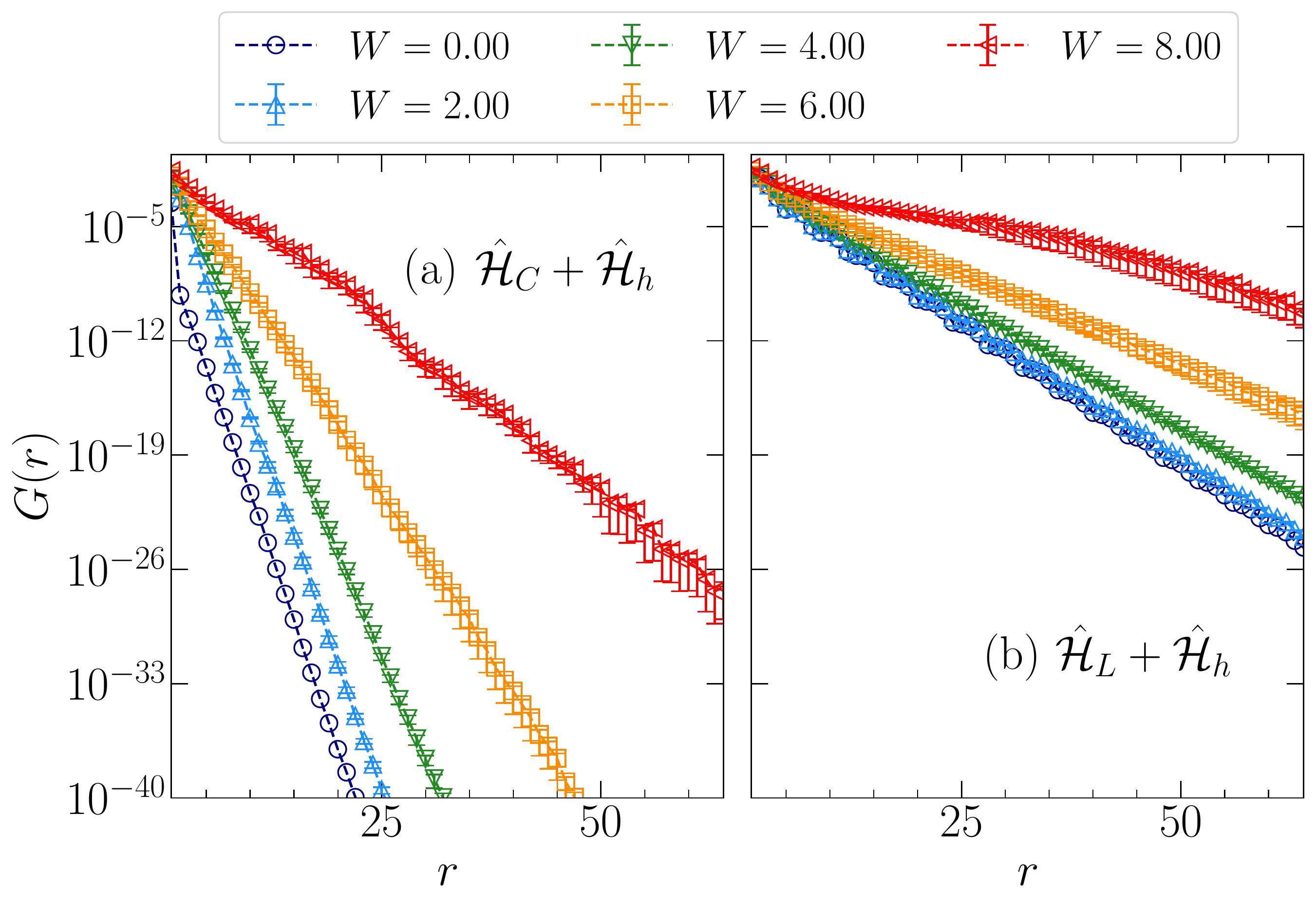}
  \caption{The same as Fig.~\ref{fig:corr_Gr_C} but for the case of disorder introduced by random Zeeman fields $\hat {\cal H}_h$. Note the robust exponential decay for a wide range of disorder amplitudes even if significantly far from the critical value $W_c$ extracted from the scaling of ${\cal D}_s$.}
  \label{fig:corr_Gr_Z}
\end{figure}

Besides of the universality class of the disorder-induced SIT in fermion-Hubbard models, there has been another long-standing question in this topic. That is, whether the route to insulating behavior proceeds through the direct localization of Cooper pairs, or by a two-step process in which the Cooper pairing is first destroyed and then followed by the standard localization of single electrons~\cite{Sacepe2011}. Alternatively, an intermediate (poor) metallic state exists where the disorder destroys the pairing coherence, but localization does not yet occur. We try to answer this question by examining the single-particle Green's function, distinguishing the metallic phase from other phases with gapped single-particle excitations, such as the superconducting state and insulating states. 

As previously mentioned, the random chemical potential does not break (local) singlet pairs. Thus the SIT transition is ought to be direct from the superconducting to the pair localization state. In that case, the single-particle Green's function would decay exponentially as the disorder strength $W$ grows, indicating that the single-particle gap remains open throughout the transition. This picture is confirmed in Fig.~\ref{fig:corr_Gr_C}, which displays the $G(r)$ decaying profile for both lattices with random chemical potentials. In contrast, the disorder introduced by the random Zeeman fields can destroy singlet pairs. However, such a disorder also destroys the coherence of pairs according to our numerical results for the superfluid weight ${\cal D}_s$. In Fig.~\ref{fig:corr_Gr_Z}, we find no clue of an intermediate metallic state with algebraic decay single-particle Green's function for both lattices with different band dispersions [see Appendix~\ref{appendix:charge_excitation} for energetic analysis of excitations supporting these results]. Finally, as an addendum, we note that a true uncorrelated Anderson insulator also exhibits gapless single-particle excitations. The fact that we observe gapped single-particle excitations across a wide range of disorder values, even substantially far from the SIT, indicates that correlation effects are still significantly relevant. If there is a transition (possibly a crossover) to such a regime, this occurs at values of $W$ where the interaction strength $|U|$ is an irrelevant perturbation.

\section{Summary and Discussion}
\label{sec:sum}

We systematically investigate the disorder-induced SIT of the attractive Hubbard model in two lattices, the Creutz lattice with noninteracting flat bands and the regular two-leg ladder with noninteracting dispersive bands. Two disorder types have been considered, random chemical potentials, which do not break local singlet pairs, and random Zeeman fields that do break pairs in general. The finite-size scaling of numerically obtained superfluid weights suggests a BKT-type phase transition for both lattices and disorder types. For the situation of non-magnetic disorder (i.e., introduced by random chemical potentials), an infinitesimal disorder drives the superconducting state to a correlated Anderson insulator of singlet pairs for both lattice geometries. For the case of the regular ladder, these results are in line with the ones obtained by a low-energy theory of the model in the regime that the attractive interactions are sufficiently large~\cite{Orignac1996, Orignac1997, Orignac1999}.

For the disorder introduced by the random Zeeman fields, the superconductivity is more robust when the noninteracting lattice has flat bands: it requires a significant disorder strength to break the superconducting state in the Creutz lattice; in contrast, the critical disorder is zero in the regular two-leg ladder. The conclusion is that the flat dispersion can enhance the superconducting state's resilience (in the presence of attractive interactions), confirmed by the pairing correlation function calculations.

One aspect that can make this comparison of results between the two types of ladder elusive is that the coordination number is not the same ($z=3$ for the regular ladder, and $z=4$ for the Creutz ladder), which can significantly impact the resilience of the superfluidity to disorder in both cases. A minimal analysis that can take this difference into account is to normalize the critical disorder $W_c$ by the total (including gaps) non-interacting bandwidth ($4t$ in the Creutz lattice and $6t$ for the regular ladder). As stated, the only case in which there is a difference between $W_c$ obtained in the Creutz and the regular ladder is when Zeeman disorder is introduced. But here, however, $W_c^{\rm C} \simeq 4.8t$ with $W_c^{\rm L} \simeq 0$, such that no (small) deviation in the band-structure widths can account for the different results observed. Nonetheless, we remark that this is certainly not the case for certain repulsive models, in which the inclusion of the extra hopping terms can fundamentally change the fate of pairing~\cite{Qin2020, Xu2023}, but mainly in a scenario of competing orders~\cite{Fradkin2015}, absent in our investigation.

Turning back to the original model, we also try to answer the long-standing question in disorder-induced SIT about whether this transition is direct or a two-step process by carefully examining the single-particle Green's function. Our results suggest no intermediate metallic state during the SIT process for all parameters involved in this work. Lastly, it is worth noting that the Hamiltonian of a Creutz ladder has already been emulated with ultracold fermionic atoms via optical potentials~\cite{Zhang2014,Kang2020}, which makes our protocol possible for experimental verification in future investigations. 

An outstanding question refers to the generality of the universality class of disorder-driven SITs in such models. While we find a clear indication of BKT-type phase transition, further supported by results in related bosonic systems~\cite{Gerster_2016, Giamarchi1988} and in the case of dispersive bands for fermions~\cite{Orignac1996, Orignac1997, Orignac1999}, this contrasts to SITs in clean systems using similar attractive Hubbard Hamiltonians~\cite{Mondaini2015, Jin2022}, which exhibit second-order phase transitions [$(d+1)$-XY universality class]. Whether this difference carries over to different dimensionalities is a question that warrants future investigation.

\section*{Acknowledgements}
We thank Edmond Orignac for pointing out pertinent references for the two-leg ladder case. C.C. was supported by the National Natural Science Foundation of China (grant nos. 11904145, 12174167, 12247101) and the Fundamental Research Funds for the Central Universities.
R.M. thanks George Batrouni and Marcos Rigol for discussions and collaborations in related contributions. R.M.~acknowledges support from NSFC Grants No.~NSAF-U2230402, 12050410263, 12111530010, 11974039, and No.~12222401. 

\appendix

\section{Single-particle excitations: clean case} \label{appendix:A_k}
In the main text, we contrast the pair excitation spectrum of both the Creutz lattice and the two-leg ladder to infer that the flatness of the Bloch bands at $U=0$ still influences the regime of strong interactions we investigate. Here we make a direct assessment, by computing the single-particle spectral function,
\begin{align}
    A_{k_x, \sigma}(\omega) &= \sum_\alpha |\langle \phi_\alpha| \hat c_{k_x,\sigma}^\dagger |\psi_0\rangle|\delta\left(\omega - (E_\alpha -E_0)\right) \nonumber \\
    &+ \sum_\alpha |\langle \phi_\alpha| \hat c_{k_x,\sigma}^{\phantom{\dagger}}|\psi_0\rangle|\delta\left(\omega + (E_\alpha -E_0)\right)\ .
    \label{eq:spectral}
\end{align}
In this expression, $|\psi_0\rangle$ is the ground-state of either \eqref{eq:Ham_Creutz} or \eqref{eq:Ham_ladder} at filling $\langle \hat n\rangle=1/4$, and the excited eigenstates states $|\phi_\alpha\rangle$ are the ones from the sectors with an added (removed) particle, $N_\sigma+1$ ($N_\sigma-1$). The operator in momentum space is defined as $\hat c_{k_x,\sigma}^\dagger = \frac{1}{\sqrt{L}}\sum_{j,y}e^{{\rm i}k_x j}\hat{c}_{j, \sigma}^{y\dagger}$, which has considered the summation of the chain index $y=A,B$.

Those are directly computed by means of a Krylov-Schur-based diagonalization method~\cite{petsc-web-page, Hernandez2005}, which we use to compute 400 eigenpairs ($E_\alpha,|\psi_\alpha\rangle$) in the low-lying spectrum of the corresponding sector, effectively truncating the summation~\eqref{eq:spectral}. Moreover, since the number of inequivalent $k$-points is $L$, we further improve the statistics by averaging each momentum value among a set of 20 equidistant twisted boundary conditions. As mentioned in the main text, these can be equivalently interpreted as threading a flux $\Phi$ on the ring ladder and can be used to mitigate finite-size effects when averaging over many values $\Phi \in [0, 2\pi)$~\cite{Poilblanc1991, Li2018}.

\begin{figure}[!h]
  \includegraphics[width=0.9\columnwidth]{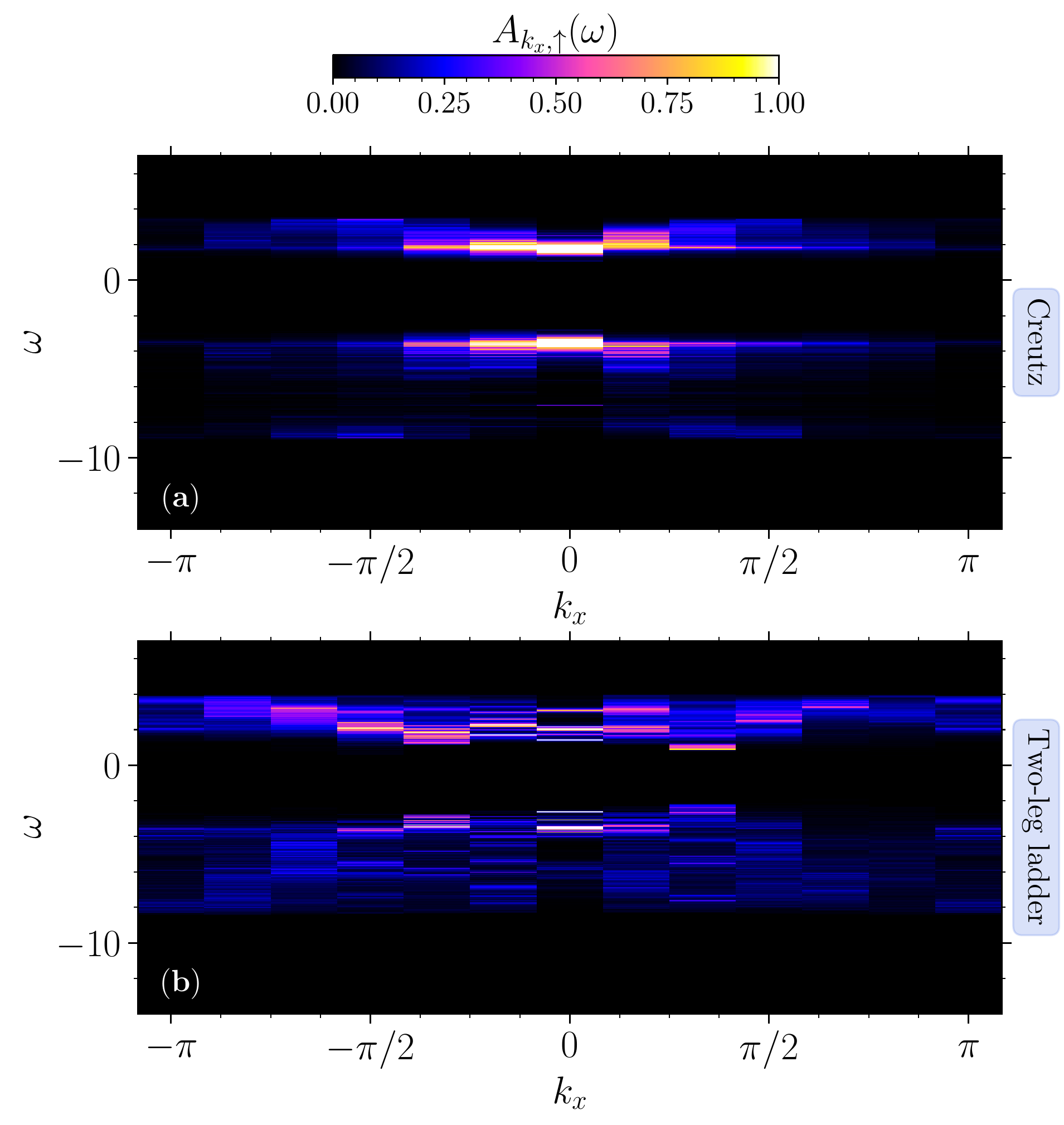}
  \caption{The heatmap of the single-particle spectral function $A_{k_x,\uparrow}(\omega)$ for the Creutz (a) and the two-leg ladder (b), in an $L=12$ lattice with Hubbard interaction $U=-8$ without disorder ($W=0$). A smaller number of dispersive features in (a) is indicative of the influence of flat-band physics in this case, even for substantially large interactions.}
  \label{fig:spectral}
\end{figure}

We report in Fig.~\ref{fig:spectral} the result for this quantity in the clean case ($W=0$), contrasting both Hamiltonians at $U=-8$, for a lattice with $L=12$, and taking into account $\sigma = \uparrow$ [$\sigma = \downarrow$ results are the same owing to the SU(2) symmetry]. A superconducting gap is clearly seen for both cases, while significant broadening of the bands occurs due to the presence of the interactions. We notice, however, that significant weight is accumulated around the $k_x=0$ excitation momentum with a flatter momentum dispersion for the Creutz ladder in comparison to the two-leg ladders, indicating the influence of the flat-band physics in this Hamiltonian type even with substantial interactions. Larger lattices (finer resolution in momentum) can potentially improve this contrast.

\section{Benchmark of DMRG results}
\label{appendix:benchmark}

\begin{figure}[!t] 
  \includegraphics[width=0.9 \columnwidth]{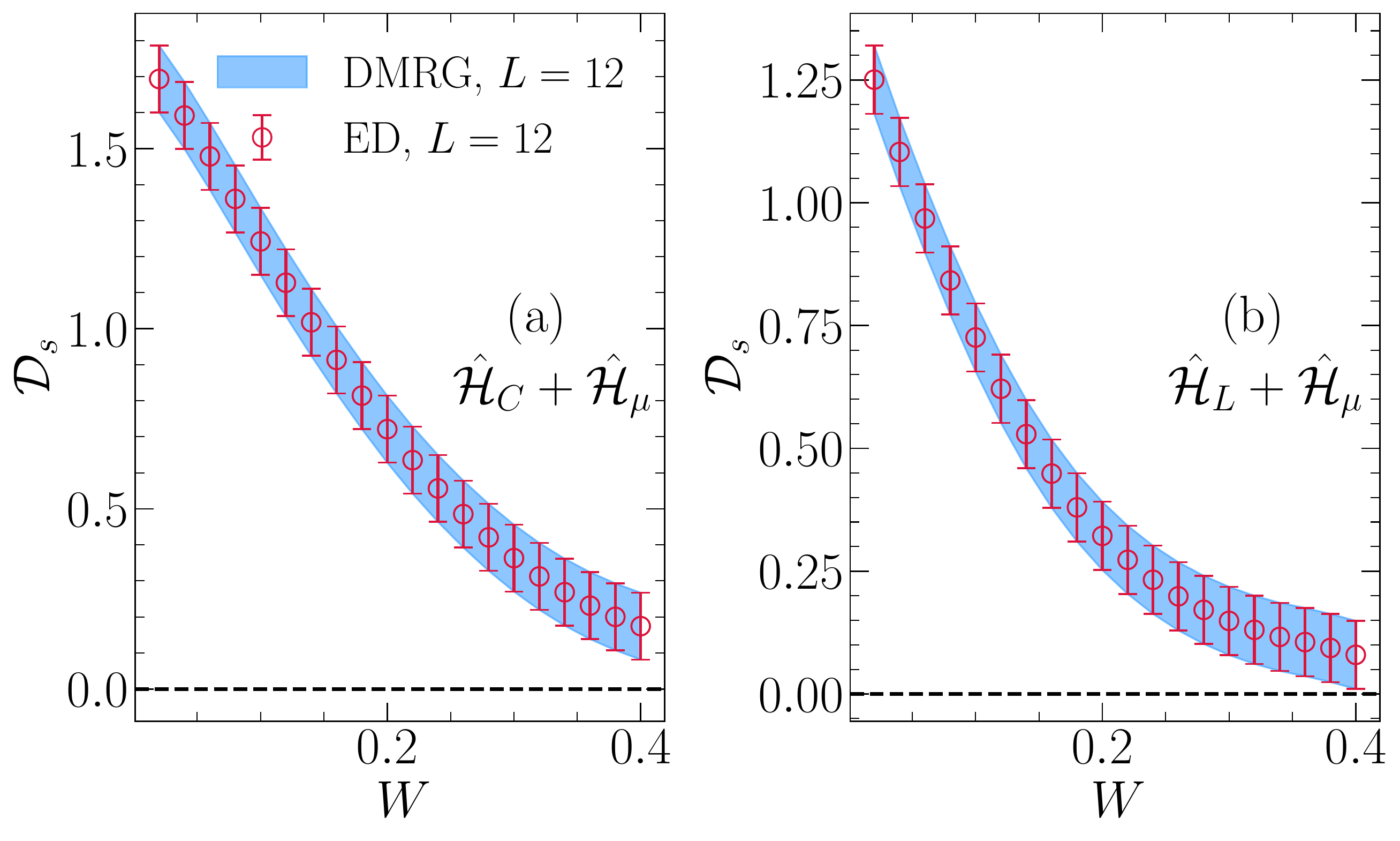}
  \caption{Comparison between ED and DMRG results for (a) Creutz lattice and (b) regular two-leg ladder in the presence of random chemical potentials. Here we use 30 disorder realizations for the benchmark. 
  }
  \label{fig:benchmark}
\end{figure}

The arguments in this work are mainly based on numerical calculations using DMRG, which is one of the most powerful methods in solving quantum many-body systems, especially in 1D and quasi-1D quantum lattices. However, in the case of periodic boundary conditions, which is precisely the case when computing the superfluid weights, DMRG meets much larger truncation errors. In other words, achieving the same precision of calculations with open boundary conditions takes a much more expensive computational effort. Moreover, when the strong disorder breaks the lattice homogeneity, the DMRG procedure is likely to be trapped in local minima, even if using an optimized strategy specially designed for disordered lattices~\cite{Xavier2018}. These difficulties, accompanied by the fact that extracting information from disordered systems requires repeating calculations for various disorder samples, restrict our investigations to relatively small system sizes. To be more rigorous, we also perform exact diagonalization (ED) calculations as a benchmark. As shown in Fig.~\ref{fig:benchmark}, the two methods provide precisely the same results, therefore confirming the reliability of the numerical results illustrated in this work. 

\section{Approximation of ${\cal D}_s$}
\label{appendix:Ds}

\begin{figure}[!t] 
  \includegraphics[width=0.9 \columnwidth]{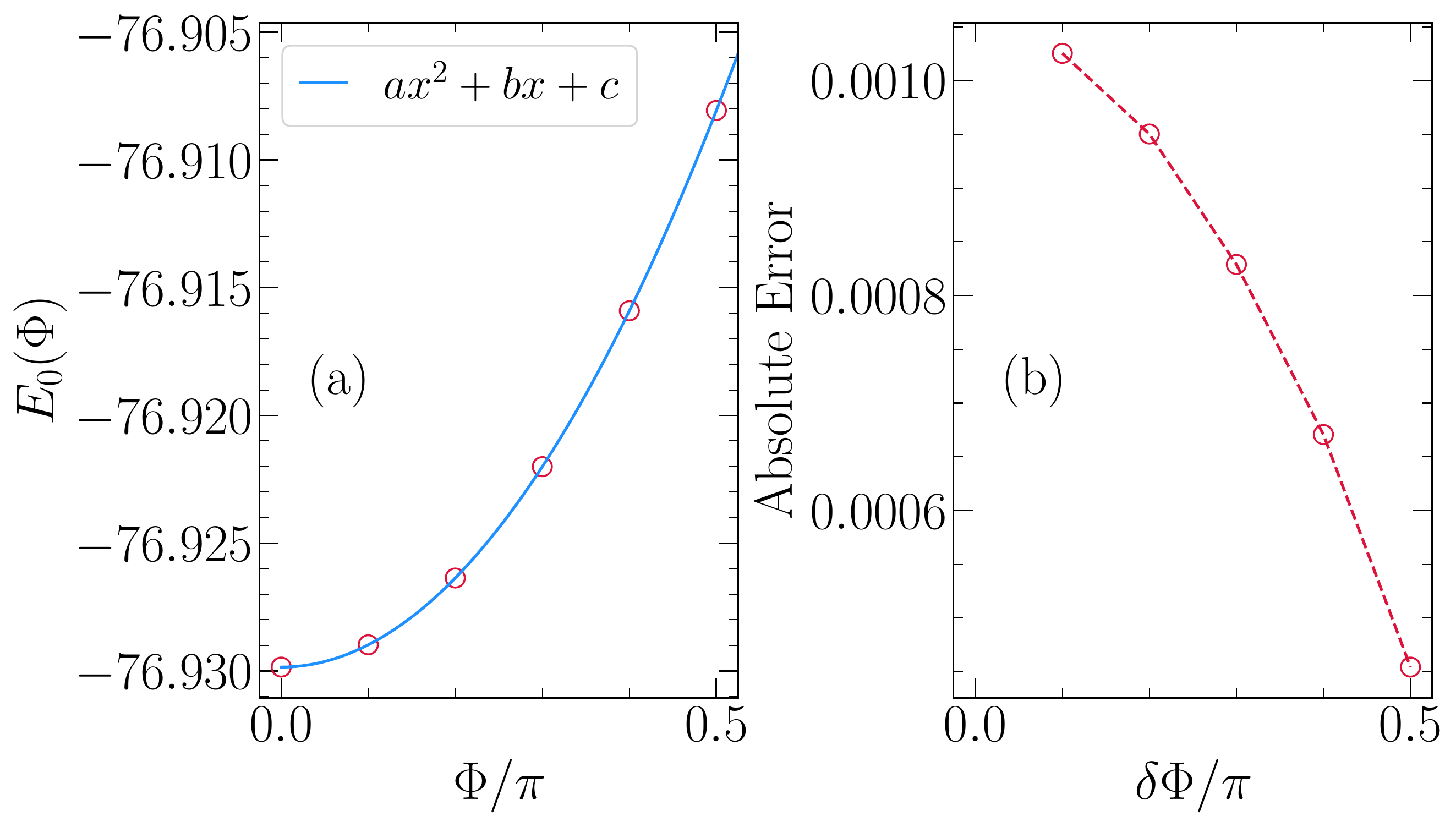}
  \caption{(a) The ground state energy $E_0(\Phi)$ versus the twisted angle $\Phi$ for the Creutz lattice in the clean case. The solid blue line denotes a second-order polynomial fitting. (b) The absolute error between the superfluid weight $D_s$ obtained from Eq.~\ref{eq:Ds} and the approximation ${\cal D}_s \approx 2\pi L [E_0(\delta\Phi)-E_0(0)]/(\delta \Phi)^2$ using different $\delta\Phi$ in Eq.~\ref{eq:approx_Ds}. Here results are from DMRG calculation of $L=32$.}
  \label{fig:Ds_error}
\end{figure}

In the absence of disorder, the ground-state energy $E_0(\Phi)$ is a quadratic function of $\Phi$ in the range $\Phi\in \left[0,\pi/2\right]$, as shown in Fig.~\ref{fig:Ds_error}(a). Therefore, the superfluid weight ${\cal D}_s$ of the form in Eq.~\ref{eq:Ds} can be obtained by the following procedure: first, do a second-order polynomial fitting of several $E_0(\Phi)$ with different twisted $\Phi$, and then compute the ${\cal D}_s$ by the second-order derivative of the previously obtained polynomial. However, this procedure is rather time-consuming, especially in the disorder case, which requires many disorder realizations. 

In practice, we adopt the approximation ${\cal D}_s \approx 2\pi L [E_0(\delta\Phi)-E_0(0)]/(\delta \Phi)^2$~\cite{Laflorencie2004}, from which one can extract $D_s$ from a single value of $E_0(\Phi)$. We display the absolute error from these two procedures in Fig.~\ref{fig:Ds_error}(b), where the error is overall small ($\sim 10^{-3}$) and decreases as the phase twist $\Phi$ increases to $\pi/2$. In this work, we choose $\Phi=\pi/2$ and use the approximation in Eq.~\ref{eq:approx_Ds} to compute the superfluid weight ${\cal D}_s$. Note that the above test has been done in the clean case, and the situation can be more complicated in the presence of a finite disorder strength. As long as $E_0(\Phi)$ is monotonic in the range $[0,\pi/2]$, the extracted ${\cal D}_s$ still likely constitutes a good approximation. Nevertheless, the results from the approximation appear promising and self-consistent in our investigation.

\begin{figure}[!b] 
  \includegraphics[width=0.9 \columnwidth]{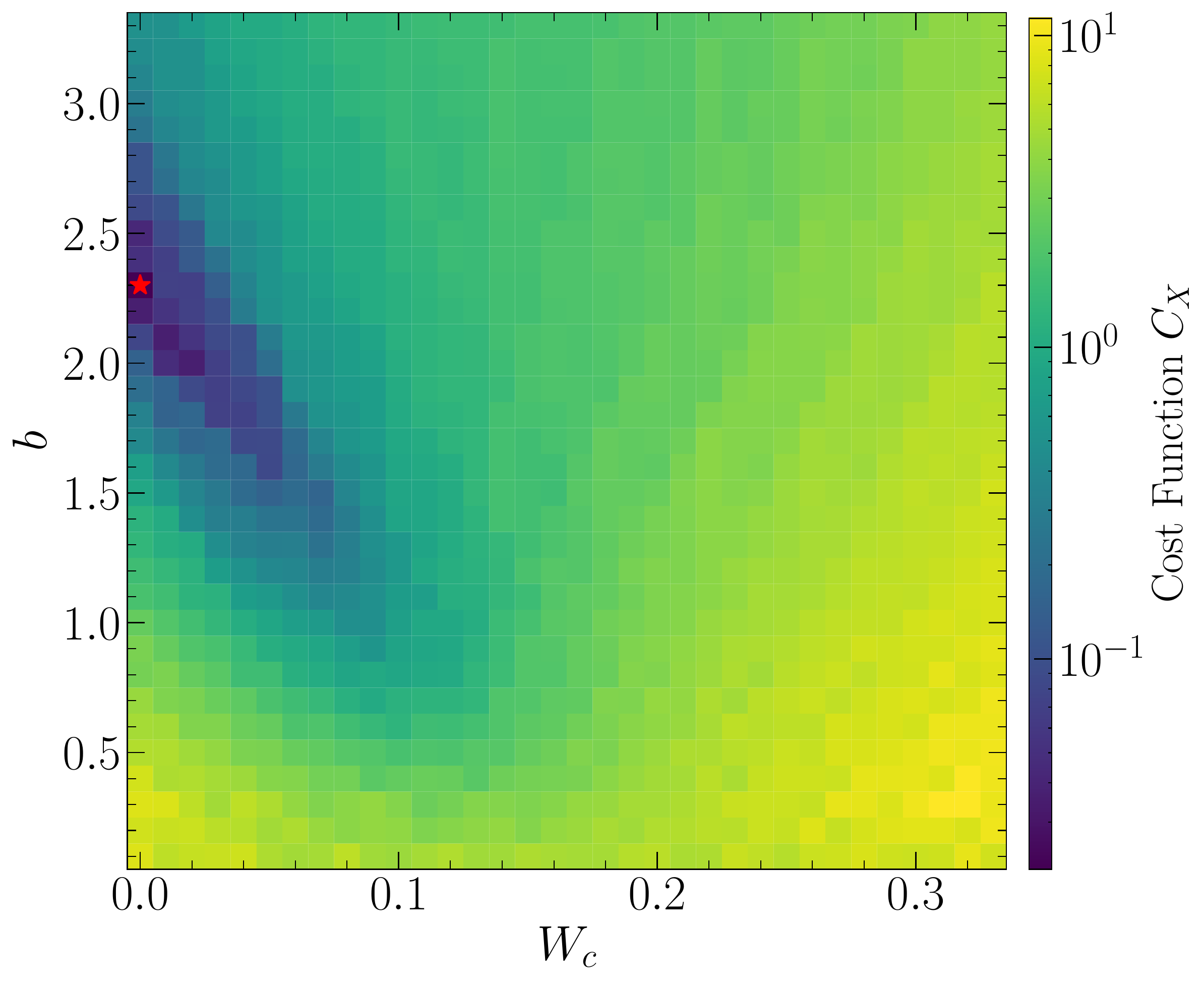}
  \caption{The cost function $C_X$ in the two-dimensional parameter space $\{ b, W_c \}$ for the Creutz lattice with random chemical potentials as an example. The red star marks the position ($W_c=0$, $b=2.3$) of the minimum $C_X$.
  }
  \label{fig:Cx_CC}
\end{figure}

\section{Cost Function Minimization}
\label{appendix:Cx}

The key for obtaining a performant data collapse and scaling of the superfluid weight is to extract the best critical $W_c$ and $b$ in Eq.~(\ref{eq:xi}), which can be determined by minimizing the cost function~\cite{Jan2020,Aramthottil2021,Mondaini2022b}
\begin{align}
    C_X=\frac{\sum_j |X_{j+1}-X_j|}{\max\{X_j\}-\min\{X_j\}}-1,
\end{align}
where $X_j$ is the $j$-th element of the collection for all $D_s(L,W)$ values in the parameter space $\{L,W\}$. Here the data collection $X$ has been sorted in a nondecreasing way with $X_j\leq X_{j+1}$. The cost function $C_X$ is close to zero for a perfectly smooth and continuous data collection. In practice, for each pair of fitted parameters value, one obtains a parameter-dependent cost function $C_X(b,W_c)$. Repeating this procedure within proper ranges in the two-dimensional parameter space $\{ b,W_c \}$ one can extract the minimum of $C_X$ and find the best fitting. As shown in Fig.~\ref{fig:Cx_CC}, the cost function of the Creutz lattice with random chemical potentials is a unimodal function in $\{ b,W_c \}$. Therefore, it is not hard to obtain the unambiguous minimum of $C_X$, and the corresponding data collapse in Fig.~\ref{fig:scaling_C} in the main text. Similar analysis carries over for the other lattice geometry and disorder type used.

\section{One- and two-particle excitation gaps}
\label{appendix:charge_excitation}

\begin{figure}[!b] 
  \includegraphics[width=0.9 \columnwidth]{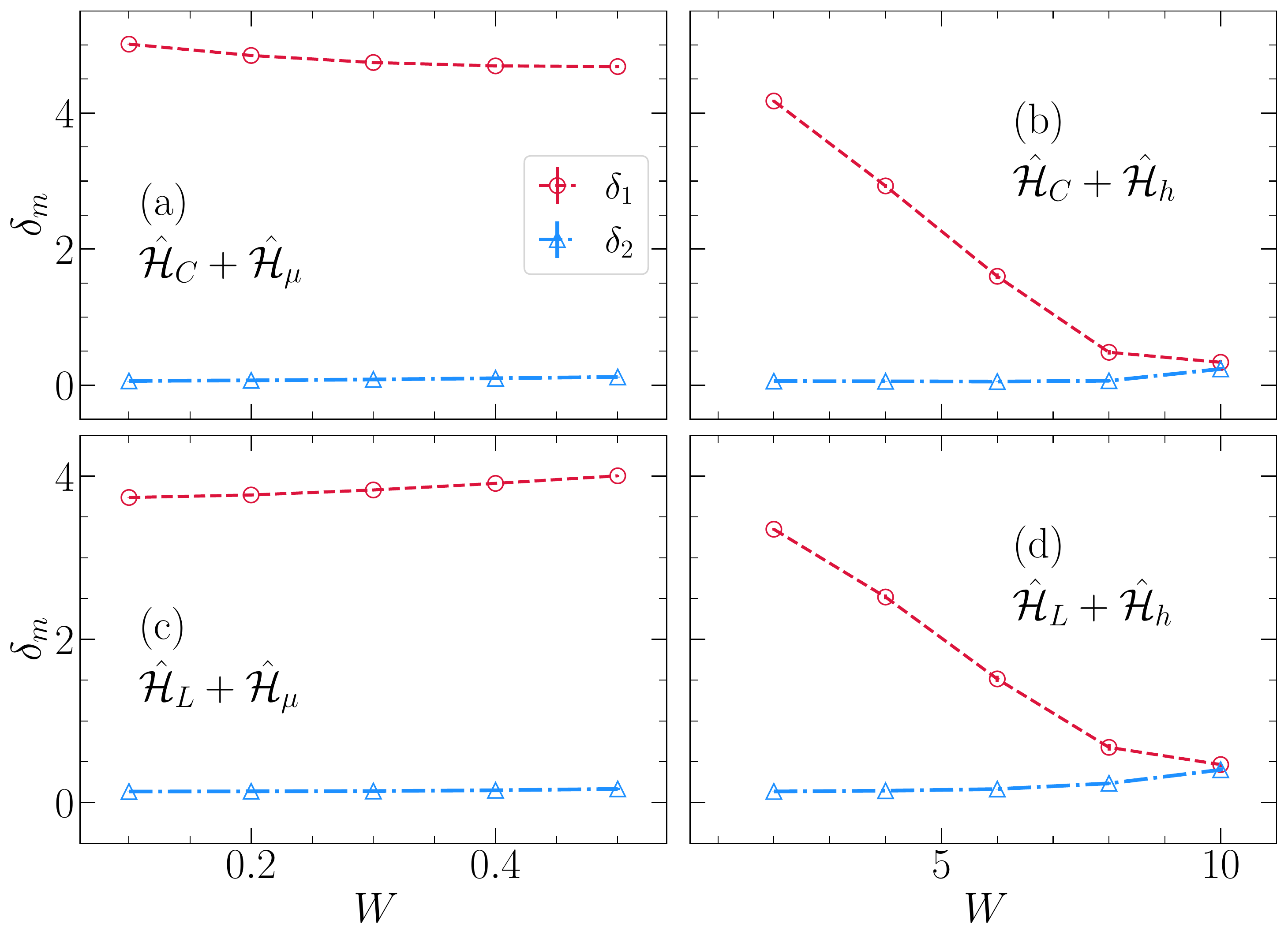}
  \caption{The $m$-particle excitation gaps $\delta_m$ (see text for definition) for (a) [(b)] Creutz lattice with random chemical potentials [Zeeman fields] and (c) [(d)] regular two-leg ladder with random chemical potentials [Zeeman fields]. Here results are from DMRG calculations with $L=12$.}
  \label{fig:E_x}
\end{figure}

On top of the observables discussed in the main text, further characterization of the different phases across the SIT can be made by examining the charge excitation energy~\cite{Sherman2014,Hazra2020,Jin2022}. In particular, the $m$-particle excitation gap can be defined as~\cite{Jin2022}
\begin{align}
    \delta_m \equiv E_0(N+m)+E_0(N-m)-2E_0(N) \ .
\end{align}
Here $E_0(N)$ is the ground state of $N=N_\uparrow+N_\downarrow$ particles, as defined in the main text. Our interest in the present work lies in the spin-balanced sector \{$N_\uparrow=N_\downarrow$\} for the case of pair excitations. In actual calculations, ($N\pm1$) [($N\pm2$)] is explicitly regarded as ($N_\uparrow\pm1,N_\downarrow$) [($N_\uparrow\pm1,N_\downarrow\pm1$)]. The one- and two-particle excitations of a small system size with $L=12$ are displayed in Fig.~\ref{fig:E_x}.

With disorder induced by random chemical potentials [Figs.~\ref{fig:E_x}(a) and \ref{fig:E_x}(c)], the one-particle excitation gap $\delta_1$ is finite in the whole range of disorder strengths investigated, irrespective of the lattice geometry (or, equivalently, the band structure). On the other hand, the two-particle excitation gap $\delta_2$ slowly grows with $W$, denoting the onset of insulating behavior. Remarkably, since $\delta_2<\delta_1$, pair excitations are favored within this regime. In the main text, we refer to it as an Anderson insulating phase of singlet pairs; in other contexts, this is also dubbed as a Bose-insulator~\cite{Nikolic2011, Nikolic2011b,Sherman2014,Hazra2020,Mondaini2015,Jin2022}. In passing, we note that this analysis also makes clear the inexistence of an intermediate disorder-induced metallic phase. 

Such scenario changes in the presence of the disorder induced by the random Zeeman fields [Figs.~\ref{fig:E_x}(b) and \ref{fig:E_x}(d)]. Now, the single-particle (two-particle) excitation gap substantially decreases (slightly increases) as $W$ grows. While having both quantities finite is a precondition for driving insulating behavior, at disorder values $W\gtrsim10$ the imminent crossing of $\delta_2$ and $\delta_1$ marks the crossover from a Bose to Fermi insulator~\cite{Mondaini2015, Hazra2020, Jin2022}, where single-particle excitations are favored instead. This change of character of the insulator phase has been seen in other contexts for clean SITs~\cite{Mondaini2015, Jin2022}. While finite-size effects likely quantitatively impact the results, they support the main findings in the main text. 

\bibliography{Creutz_lattice_refs}
\end{document}